\documentclass[aps,onecolumn,pra,superscriptaddress,showpacs,showkeys,floatfix]{revtex4-1}
\usepackage{graphicx,color}
\usepackage{epsfig}
\usepackage{graphicx}
\usepackage{mathrsfs}
\usepackage{color}
\usepackage{bm}
\usepackage{amsmath,amssymb,amsthm,amscd}
\usepackage{bbold}
\usepackage{mdwlist}
\newcommand{\proj}[1]{|#1\rangle\langle#1|}

 \newcommand{\bra}[1]{\langle{#1} |}
 \newcommand{\ket}[1]{|{#1}\rangle  }
 
 \newcommand{\ketbra}[2]{\vert {#1} \rangle \langle{#2}\vert}

\newcommand{\Tr}{\operatorname{Tr}}

\begin{document}

\title{Detection of properties and capacities of quantum channels}

\author{Chiara Macchiavello}
\affiliation{Quit group, Dipartimento di Fisica, 
Universit\`a di Pavia, via A. Bassi 6, 
 I-27100 Pavia, Italy}
\affiliation{Istituto Nazionale di Fisica Nucleare, Sezione di Pavia, via A. Bassi 6,
  I-27100 Pavia, Italy}

\author{Massimiliano F. Sacchi}
\affiliation{Istituto di Fotonica e Nanotecnologie - CNR, Piazza Leonardo
  da Vinci 32, I-20133, Milano, Italy}
\affiliation{Quit group, Dipartimento di Fisica, 
Universit\`a di Pavia, via A. Bassi 6, 
 I-27100 Pavia, Italy}
	
\date{\today}

\begin{abstract} 

We review in a unified way a recently proposed method to detect
properties of unknown quantum channels and lower bounds to quantum
capacities, without resorting to full quantum process tomography. The
method is based on the preparation of a fixed bipartite entangled
state at the channel input or, equivalently, an ensemble of an
overcomplete set of single-system states, along with few local
measurements at the channel output.

\end{abstract}

\maketitle

\section{Introduction}

Noise is unavoidably present in any communication channel and it affects the 
efficiency with which quantum states can be transmitted. 
A complete characterisation of a quantum channel can be achieved by means
of quantum process tomography \cite{nielsen97,pcz,mls,dlp,alt,cnot,vibr,
ion,mohseni,irene, atom},  but this procedure requires a large number of 
measurement settings when it has to be implemented experimentally (which scales  
as $d^4$, where $d$ is the dimension of the quantum system on which the 
channel acts). In many practical situations a complete characterisation 
is not needed and only some features of a quantum channel may be of interest.
It is threfore of great importance to achieve information about specific
properties of a quantum channel, and more generally of the quantum evolution
of an open quantum system, with the minimum 
number of measurement settings. In this work we review a recently proposed 
method to achieve information on some properties of a quantum channel
 \cite{mapdet} and on the channel ability of transmitting quantum information
 \cite{qcap-det}
by employing a number of measurement settings that scales more favourably with 
respect to quantum process tomography, i.e. as $d^2$.

The present paper is organised as follows. In Sect. \ref{s:preli} we present 
the general picture by setting the scenario and reminding 
some preliminary notions that represent the main ingredients to develop
the proposed quantum channel detection method.
In Sect. \ref{s:mapdet} we illustrate the method to detect convexity 
properties of quantum channels, such as being or not entanglement breaking.
In Sect. \ref{s:qcap} we specify the method to detect lower bounds to
quantum channel capacities and describe it extensively in the case of qubit
channels.  In Sect. \ref{s:conc} we finally summarise the main results.

\section{General scenario}
\label{s:preli}

Quantum channels, and in general quantum noise processes that describe
the evolution of open quantum systems, 
are described by completely positive and trace preserving 
(CPT) maps ${\cal E}$, which can be expressed in the Kraus form 
as
\begin{equation}
{\cal E}(\rho )=\sum_k A_k\rho A_k^\dagger,
\label{kraus}
\end{equation}
where $\rho$ is the density operator of the quantum system on which the 
channel acts, and the Kraus operators $\{A_k\}$ fulfil the completeness constraint 
$\sum_k A_k^\dagger A_k= I$. In this paper we consider quantum channels acting 
on systems with finite dimension $d$, also referred to as qudits.

In order to develop the detection method proposed, we will use the 
Choi-Jamiolkowski isomorphism \cite{jam, choi}, which gives a one-to-one
correspondence between CPT maps ${\cal E}$ acting on the set of 
density operators on Hilbert space $\mathcal{H}$ and bipartite density 
operators 
$C_{{\cal E}}$ on $\mathcal{H\otimes H}$. This isomorphism can be described as
\begin{equation}
{\cal E}\Longleftrightarrow 
C_{{\cal E}}={\cal E}\otimes {\cal I} (\ket{\phi^+}\bra{\phi^+}),
\end{equation}
where ${\cal I}$ is the identity map, 
and $\ket{\phi^+}$ is the maximally entangled state with respect to the 
bipartite space $\mathcal{H\otimes H}$, i.e. $\ket{\phi^+}=\frac{1}{\sqrt 
d}\sum_{k=0}^{d-1}\ket{k}\ket{k}$. 

By exploiting the above isomorphism, we are able to link some specific 
properties of quantum channels to properties of the corresponding 
Choi-Jamiolkowski states 
$C_{{\cal E}}$, as we will review in Sect.
\ref{s:mapdet}. In particular, a connection between quantum channel 
properties and entanglement properties of the corresponding 
Choi-Jamiolkowski states can be established. 
The method works when we consider properties 
that are based on a convex structure of the quantum channels. 

We first define a basis of maximally entangled states for bipartite
$d$-dimensional systems as 
\begin{eqnarray}
|\Phi^{mn} \rangle = (U_{mn}\otimes I ) |\phi^+ \rangle \;,
\qquad m,n=0,1,\cdots ,d-1 \,,\label{bellui}
\end{eqnarray}
where $U_{mn}$ represents the unitary operator 
$U_{mn}=\sum _{k=0}^{d-1} e^{\frac{2\pi i}{d} km} |k \rangle 
\langle (k + n)\!\!\!\mod d |$, satisfying the orthogonality relations 
$\Tr [U_{mn} ^\dag  U_{pk}]= d\, \delta _{mp}
\delta _{nk}$. A set of generalised Bell projectors can then be explicitly 
written as follows \cite{bellob}
\begin{eqnarray}
\ketbra{\Phi ^{mn}}{\Phi ^{mn}}=\frac {1} {d^2 }\sum _{p,q=0} ^{d-1} 
e^{\frac {2\pi i}{d}(np-mp) }U_{pq} \otimes U^*_{pq} 
\,,\label{umnbel}
\end{eqnarray}
where $m,n=0,1,\cdots, d-1$, and $*$ denotes complex conjugation. 
Notice that $U_{00}$ corresponds to the 
identity operator and $|\Phi^{00} \rangle \equiv |\phi^+ \rangle$.

The scenario we will focus on to achieve our detection strategy 
consists in the following steps: prepare a 
bipartite pure 
state $|\phi^+\rangle $ composed of a system qudit and a 
noiseless reference qudit denoted by $R$, and send it through the channel 
${\cal E}\otimes{\cal I} _R$, where the unknown channel ${\cal E}$ acts on the
system qudit. Then measure only the $d^2-1$ local observables  
$ {\cal M}_d= \{U_{mn} \otimes U^*_{mn}\,, \, m,n=0,1,\cdots ,d-1 \,; m+n\neq 0\}$ 
 on the system and on the reference qudits.
This is the basic scenario that pertains to the detection methods that will
be outlined in the following sections.

We want to point out here that such methods based on the measurements of the 
local operators 
$U_{mn} \otimes U^*_{mn}$ do not necessarily require the use of an entangled 
bipartite state at the input. Actually, the same results can be achieved 
by considering the single system qudit on which the channel acts as follows. 
Using the identity \cite{bellob} 

\begin{equation}
\langle \phi^+| B \otimes C |\phi^+\rangle =\frac{1}{d}\Tr [B C^\tau]\;,
\end{equation} 
valid for any pair of Hilbert-Schmidt operators $B$ and $C$ acting
on $\mathcal{H}$, where $C^\tau$ denotes the transposed operator, 
and remembering the Kraus form reported in Eq. (\ref{kraus}), we can write
the following identity for the expectations on the bipartite output state  
\begin{equation}
\langle U_{mn}  \otimes U^*_{pk} \rangle =
\Tr [({\cal E}\otimes{\cal I} _R)(|\phi ^+ \rangle \langle \phi ^+ |) 
( U_{mn}  \otimes U^*_{pk} )]=\frac 1 d 
\Tr[U_{mn} {\cal E}(U_{pk}^{\dag })]\;.
\end{equation} 
The expectation values $\langle U_{mn}  \otimes U^*_{mn} \rangle $ can then be 
obtained by considering only the system qudit, preparing it in the 
eigenstates of $U_{mn}^{\dag }$ with equal probabilities, and 
measuring $U_{mn}$ at the output of the channel.

\section{Detection of convex properties}
\label{s:mapdet}

We will now show how to detect convex properties of quantum channels
that may be related to entanglement properties of the Choi-Jamiolkowski state
following the method originally proposed in Ref. \cite{mapdet}.
The main ingredient that is employed for this purpose is the concept of
entanglement detection via witness operators \cite{horo-ter}. 
We remind here that a state $\rho$ is entangled 
if and only if there exists a hermitian operator $W$ such that 
$\Tr[W\rho]< 0$ and $\Tr[W\rho_{sep}]\geq 0$ for all separable states.
We illustrate explicitly the channel detection procedure by considering 
unknown qubit channels 
and asking whether a given channel is entanglement breaking.
A possible definition for an entanglement breaking channel is
based on the separability of its Choi-Jamiolkowski state: 
a quantum channel is entanglement breaking if and only if its 
Choi-Jamiolkowski state is separable.
The set of entanglement breaking channels is a convex set and, clearly, 
the set of Choi-Jamiolkowski states corresponding to entanglement breaking 
channels contains only bipartite separable states.
This allows to formulate a method to detect whether a quantum channel is not
entanglement breaking by exploiting entanglement detection methods designed
for bipartite systems \cite{ent-wit}.

As a simple example of quantum channel detection consider the case of 
qudits and a generalised depolarising channel. This is described by
a particular case of the generalised Pauli channel defined as follows
\begin{equation}\label{dep-channel}
{\cal E}(\rho)= \sum^{d-1}_{m,n=0}{p_{m,n}U_{mn} \rho U_{mn}^*}\;,
\end{equation}
where $p_{m,n}$ are probabilities.
In the depolarising case we have a special form for the probabilities, namely
$p_{0,0}=1-p$ (with $p\in[0,1]$), while $p_{m,n}=p/(d^2-1)$ for $m,n=0,\cdots ,d-1$
with $m+n\neq 0$. Such a channel is entanglement breaking for $p \geq 1/(d+1)$. 
The corresponding set of Choi-Jamiolkowski bipartite density operators 
is given by the Werner states
\begin{equation}\label{werner states}
\rho_p= \left( 1-\frac{d^2}{d^2-1}p \right ) \proj{\phi^+}
+\frac{p}{d^2-1} I \otimes I_R \;.
\end{equation}
The above states are entangled for $p\leq (d-1)/d$ \cite{pitt-rubin}. 
It is then possible to detect whether a depolarising channel is not 
entanglement 
breaking by exploiting an entanglement witness operator for the above set
of states \cite{ent-wit,jmo}, which has the form
\begin{equation}\label{wEB}
W_{EB,d}= \frac{d-1}{d^2}I \otimes I_R  
-\frac{1}{d^2}\sum _ {m+n >0}
U_{mn} \otimes U^*_{mn}\;.
\end{equation}

The method can then be implemented by employing the scheme outlined
in the previous section and evaluating $\Tr[W_{EB}\rho_p]$ by 
the $d^2-1$ local measurements performed on the two qudits. If the resulting 
average value is negative, we can then conclude that the channel under 
consideration is not entanglement breaking.

In the particular case of two-dimensional systems the above scenario 
corresponds to the measurement of local Pauli operators 
$ {\cal M}_2= \{\sigma _x \otimes \sigma _x , 
\sigma _y \otimes \sigma _y , \sigma _z 
\otimes \sigma _z \}$  and a suitable operator to detect non entanglement 
breaking channels takes the simple form
\begin{equation}\label{wEB2}
W_{EB,2}= \frac{1}{4}(I\otimes I_R 
-\sigma_x\otimes \sigma_x +\sigma_y\otimes \sigma_y -\sigma_z\otimes 
\sigma_z)\;.
\end{equation}

We point out that the scheme outlined here can be generalised to detect other 
properties of quntum 
channels based on convexity features, such as those related to being a 
separable random unitary channel, separable or PPT channels \cite{mapdet},
and completely co-positive or bi-entangling operations \cite{conf-phys-ts}.
We want to stress that the scheme is very simple to implement in an 
experimental scenario and it was successfully tested in Ref. \cite{exp-qcd}
with single qubit and two-qubit channels.

\section{Detection of quantum capacities}
\label{s:qcap}

In this Section we address the situation where we want to certify the ability
of a channel to transmit quantum information by avoiding the use of full 
quantum process tomography. Our purpose is to  
employ a smaller number of measurements, that for arbitrary finite
dimension $d$ scales as $d^2$ as addressed in the previous sections.
We will first consider the case of detection of the quantum capacity,
following the approach developed in \cite{qcap-det}.

\par In the following we focus on memoryless channels. 
As above, we denote the action of a 
generic quantum channel on a single system as ${\cal E}$ and define 
${\cal E}_N= {\cal E}^{\otimes N}$, where $N$ represents the number of 
channel uses.
The quantum capacity $Q$ measured in qubits per channel use is defined 
as \cite{lloyd,barnum,devetak}
\begin{eqnarray} Q=\lim _{N\to \infty}\frac
{Q_N}{N}\;,\label{qn} 
\end{eqnarray} 
where
$Q_N = \max
_{\rho } I_c (\rho , {\cal E}_N)$, 
and $I_c(\rho , {\cal E}_N)$ is the coherent information 
\cite{schumachernielsen}
\begin{eqnarray} I_c(\rho , {\cal E}_N) = S[{\cal E}_N (\rho )] - S_e
(\rho, {\cal E}_N)\;.\label{ic} \end{eqnarray} 
In Eq. (\ref{ic})
we denote with $S(\rho )=-\Tr [\rho \log _2 \rho ]$ 
the von Neumann entropy, and
with $S_e (\rho, {\cal E})$  the entropy exchange \cite{schumacher}, i.e.
$S_e (\rho, {\cal E})= S[({\cal E} \otimes {\cal I}_R 
)(|\Psi _\rho \rangle \langle \Psi _\rho |)] $,
where $|\Psi _\rho \rangle $ is any purification of $\rho $ by means of a 
reference quantum system $R$, namely 
$\rho =\Tr _R [|\Psi _\rho \rangle \langle \Psi _\rho|]$.

\par We now briefly review the derivation of the lower bound of Ref.
\cite{qcap-det} for the quantum 
capacity $Q$ that can be easily accessed without requiring full process 
tomography of the quantum channel. Since for any complete set of orthogonal 
projectors $\{\Pi _i\}$ one has \cite{NC00} 
$S(\rho )\leq  S(\sum _i \Pi _i \rho \Pi _i)$, then for any orthonormal basis 
$\{ |\Phi _i \rangle \}$ in the tensor product of system and reference 
Hilbert spaces one has the following upper bound to the entropy exchange
\begin{eqnarray}
S_e\left (\rho , {\cal E} \right )\leq H (\vec p)\;,  
\label{se-bound}
\end{eqnarray}
where $H(\vec p)$ denotes the Shannon entropy for the vector of the 
probabilities $\{p_i\}$,  with 
\begin{eqnarray}
p_i = \Tr [( {\cal
E}\otimes {\cal I}_R)
(|\Psi _\rho \rangle \langle \Psi _\rho |) |\Phi _i \rangle\langle\Phi _i|] 
\;.
\label{pimeas}
\end{eqnarray}

From Eq. (\ref{se-bound}) it follows that for any input density operator 
$\rho$ and vector of probabilities $\vec p$ one has 
the following chain of bounds 
\begin{eqnarray}
Q \geq Q_1 \geq I_c(\rho , {\cal E}_1)\geq S\left [{\cal E} (\rho )\right ]-H(\vec p) 
\equiv Q_{DET}
\;.\label{qvec}
\end{eqnarray}
A lower bound $Q_{DET}$ to the quantum capacity of an unknown channel can 
then be detected  by using the scheme described in Sect. \ref{s:preli}, 
where a bipartite pure 
state $|\phi^+ \rangle $ is prepared and sent through the channel 
${\cal E}\otimes {\cal I} _R$. The set of local observables $ {\cal M}_d$
is then measured on the joint output state: in this way it is possible  
to estimate 
$\vec p$ and $S\left [{\cal E} (\rho )\right ]$ in order to compute $Q_{DET}$. 
Notice that
for the system alone the considered observables correspond to a 
tomographically complete set 
of measurements and they allow to perform an exact estimate of the term
$S\left [{\cal E} (\rho )\right ]$ in Eq. (\ref{qvec}).
Moreover, in principle, in a more general scenario, one can even adopt an 
adaptive detection scheme to 
improve the bound (\ref{qvec}) by varying the input state 
$|\Psi _\rho \rangle $.

\par We will now illustrate the efficiency of the method by considering 
some specific forms of quantum channels. We will start from
the depolarising channel  
in arbitrary dimension $d$, already introduced in the previous section,
whose action of 
(\ref{dep-channel}) can also be written as
\begin{eqnarray}
{\cal E}(\rho )=
\left ( 1 - p \frac {d^2}{d^2-1}  \right )\rho  + p 
\frac {d^2}{d^2 -1}  \frac Id \;. 
\end{eqnarray}
In this case the detectable bound is simply given by 
\begin{eqnarray}
Q \geq Q_{DET}=\log _2 d - H_2 (p) - p\log _2 (d^2 -1)\;,\label{hashd}
\end{eqnarray}
where $H_2(x) \equiv  -x \log_2 x -(1-x)\log _2 (1-x)$ denotes the binary Shannon entropy, 
and can be detected by estimating $\vec p$ pertaining to the Bell projectors 
(\ref{umnbel}). 

\par As mentioned above, this noise model can be generalised to a generic 
Pauli channel of the form (\ref{dep-channel}).
In this case the detectable bound is generalised to
\begin{eqnarray}
Q \geq Q_{DET}=\log _2 d - H(\vec p) \;, 
\end{eqnarray}
where $\vec p$ is now the $d^2$-dimensional vector of probabilities $p_{mn}$ 
pertaining to the  
generalised Bell projectors in Eq. (\ref{umnbel}). We notice that our detectable bound coincides with the 
theoretical hashing bound \cite{hashing}.

We will now focus on the specific case of qubit channels. 
By explicitly denoting the Bell states as
\begin{eqnarray}
&&|\Phi ^\pm \rangle =\frac {1}{\sqrt 2}
(|00 \rangle \pm |11 \rangle )\,, 
\ \  |\Psi ^\pm \rangle =\frac {1}{\sqrt 2}
(|01 \rangle \pm |10 \rangle )\,, 
\label{phipsi}
\end{eqnarray}
it can be straightforwardly proved that the local measurement settings 
$ {\cal M}_2$ allow to estimate the vector $\vec p$ pertaining to 
the projectors onto the following inequivalent bases
\begin{eqnarray}
B_1= &&\{ a |\Phi ^+ \rangle + b |\Phi ^- \rangle  , 
-b |\Phi ^+ \rangle + a |\Phi ^- \rangle  ,  
c |\Psi ^+ \rangle + d |\Psi ^- \rangle  , -d  |\Psi ^+ \rangle + c |\Psi ^- \rangle   
\}\;,
\label{b1} 
\\ 
B_2= &&\{ a |\Phi ^+ \rangle + b |\Psi ^+ \rangle  , 
-b |\Phi ^+ \rangle + a |\Psi ^+ \rangle  , 
c |\Phi ^- \rangle + d |\Psi ^- \rangle  , -d  |\Phi ^- \rangle + c |\Psi ^- \rangle  
\}
\;,\label{b2} 
\\ 
B_3= &&\{ a |\Phi ^+ \rangle + i b |\Psi ^- \rangle  , 
i b |\Phi ^+ \rangle + a |\Psi ^- \rangle  , 
c |\Phi ^- \rangle + i d |\Psi ^+ \rangle  , i d  |\Phi ^- \rangle + c |\Psi ^+ \rangle   
\}
\;,\label{b3} 
\end{eqnarray}
with $a,b,c,d$ real and such that $a^2+b^2=c^2+d^2=1$. 
Actually, the measurements corresponding to the above three bases are 
achieved by orthogonal projectors of the form
\begin{eqnarray} 
&&\Pi _{\{a | \Phi ^+\rangle + b| \Phi ^-\rangle\}}=\frac 14 (I\otimes I_R + 
\sigma _z \otimes \sigma _z  )  
+\frac {a^2 -b^2}{4}(\sigma _x \otimes \sigma _x - \sigma _y \otimes \sigma _y ) +\frac {ab}{2}(\sigma _z \otimes I_R + 
I\otimes \sigma _z )\,, 
\\ 
&&\Pi _{\{c |\Psi ^+\rangle + d |\Psi ^- \rangle \}}=\frac 14 (I\otimes I_R  - 
\sigma _z \otimes \sigma _z  )  
+\frac {c^2 -d^2}{4}(\sigma _x \otimes \sigma _x + \sigma _y \otimes \sigma _y ) +\frac {cd}{2}(\sigma _z \otimes I_R - 
I\otimes \sigma _z )\,,
\\ 
&&\Pi _{\{a |\Phi ^+ \rangle + b |\Psi ^+ \rangle \}}=\frac 14 (I\otimes I_R + 
\sigma _x \otimes \sigma _x  )  
+\frac {a^2 -b^2}{4}(\sigma _z \otimes \sigma _z - \sigma _y \otimes \sigma _y ) +\frac {ab}{2}(\sigma _x \otimes I_R + 
I\otimes \sigma _x )\,, 
\\ 
&&\Pi _{\{c |\Phi ^- \rangle + d |\Psi ^- \rangle \}}=\frac 14 (I\otimes I_R - 
\sigma _x \otimes \sigma _x  )  
+\frac {c^2 -d^2}{4}(\sigma _z \otimes \sigma _z +\sigma _y \otimes \sigma _y  ) -
\frac {cd}{2}(\sigma _x \otimes I_R - 
I\otimes \sigma _x )\,,
\\ 
&&\Pi _{\{a |\Phi ^+ \rangle + i b |\Psi ^- \rangle \}}=\frac 14 (I\otimes I_R - 
\sigma _y \otimes \sigma _y  )  
+\frac {a^2 -b^2}{4}(\sigma _z \otimes \sigma _z + \sigma _x \otimes \sigma _x  ) -\frac {ab}{2}(\sigma _y \otimes I_R - 
I\otimes \sigma _y )\,,
\\ 
&&\Pi _{\{c |\Phi ^- \rangle + i d |\Psi ^+ \rangle \}}=\frac 14 (I\otimes I_R + 
\sigma _y \otimes \sigma _y  )  
+\frac {c^2 -d^2}{4}(\sigma _z \otimes \sigma _z - \sigma _x \otimes \sigma _x ) +\frac {cd}{2}(\sigma _y \otimes I_R + 
I\otimes \sigma _y )\,, 
\end{eqnarray} 
where $\Pi _{\{a |\Phi ^+\rangle  + b | \Phi ^- \rangle\}}$ denotes the projector 
onto the state $a |\Phi ^+\rangle + b |\Phi ^- \rangle$, and analogously 
for the other projectors.
The probability vector $\vec p$ for each choice of basis 
is then evaluated according to Eq. 
(\ref{pimeas}). The expectation values for terms of the form $\sigma_x\otimes 
I_R$ (or $I\otimes \sigma _x$) 
can be measured from the outcomes of the observable  $\sigma _x \otimes \sigma _x$ by 
ignoring the measurement results on the second (or first) qubit, and analogously for 
the other similar terms in the above projectors.

Therefore, the bound $Q_{DET}$ in (\ref{qvec}) given the 
fixed local measurements $\{\sigma _x \otimes \sigma _x,\sigma _y \otimes 
\sigma _y,\sigma _z \otimes \sigma _z \}$ can be optimised if 
the Shannon entropy 
$H(\vec p)$ will be minimised as a function of the bases (\ref{b1}-\ref{b3}), 
by varying the coefficients $a,b,c,d$ over the three sets. 
In an experimental scenario, after collecting the outcomes of the measurements 
$\{\sigma _x \otimes \sigma _x,\sigma _y \otimes 
\sigma _y,\sigma _z \otimes \sigma _z \}$, this optimisation 
step corresponds to classical processing of the measurement outcomes.

The detectable bound can then be optimised as
\begin{eqnarray}
Q_{DET}=\max _{i=1,2,3}\max _{b,d}Q_{DET}(B_i,b,d) 
=S[\mathcal{E}(\rho)]-\min _{i=1,2,3}\min _{b,d}
H[\vec{p}(B_i,b,d)]\;.\label{opt}
\end{eqnarray}

We will now study some specific forms of qubit channels.  
A dephasing channel for qubits with unknown probability $p$ can be written 
as 
${\cal E}(\rho )= \left (1-\frac p 2 \right ) 
\rho + \frac p 2 \sigma _z \rho \sigma _z
$.  
Since it is a degradable channel, 
its quantum capacity coincides with the one-shot single-letter quantum 
capacity $Q_1$, and one has   
$Q=Q_1= 1 - H_2 \left( \frac p2 \right )$.
The von Neumann entropy of the output state 
${\cal E}(\frac I 2)=\frac I 2$ is given by $S\left [{\cal E}\left (\frac I 2 \right )\right ]
=1$. Using the Bell basis (\ref{phipsi}) one finds that 
the detectable bound coincides with the quantum capacity, namely  $Q_{DET}\equiv Q$. 

\par The depolarising channel with probability $p$ for qubits is given by 
 ${\cal E}(\rho )=(1-p)\rho +\frac p3 \sum _{i=x,y,z}\! \sigma _i \rho \sigma _i $.  
The quantum capacity is still unknown, although one has the upper bound \cite{ssw} 
$Q \leq 1 - 4p$, 
thus showing that $Q=0$ for $p\geq \frac 14$. 
On the other hand, by random coding the following hashing bound \cite{hashing}
has been proved
\begin{eqnarray}
Q \geq 1- H_2 (p) - p\log _2 3\;.\label{hash}
\end{eqnarray}
This lower bound coincides with our detectable bound $Q_{DET}$ by using the Bell basis in Eq. (\ref{phipsi}). 
Our procedure allows to certify $Q(p)\neq 0$ as long as $p <  0.1892 $.

\par In order to illustrate explicitly the usefulness of the classical
optimisation over the measurement results given in Eq. (\ref{opt}), we consider
the amplitude damping channel for qubits, that has the form \cite{NC00}
\begin{eqnarray}
{\cal E}(\rho )= A_0 \rho A_0^\dag + A_1  \rho A_1^\dag \;, 
\end{eqnarray}
where $A_0= |0 \rangle \langle 0| + \sqrt {1- \gamma }|1 \rangle \langle 1|$ and 
$A_1= \sqrt \gamma |0 \rangle \langle 1|$. 
Since it is a degradable channel \cite{gf}, 
its quantum capacity coincides with the one-shot 
single-letter 
quantum capacity $Q_1$, and it is given by   
\begin{eqnarray}
Q=Q_1= \max _{q\in [0,1]} \{ H_2[(1 - \gamma  )q] - H_2(\gamma q) \}
\;,\label{qcdamp}
\end{eqnarray}
for $ \gamma   \leq \frac 12$, and $Q=0$ for $ \gamma \geq \frac 12$. 
In our procedure, by starting from an input Bell state $|\Phi ^+ \rangle $, 
the output can be explicitly written as
\begin{eqnarray}
&&{\cal E} \otimes {\cal I}_R
(|\Phi ^+ \rangle \langle \Phi ^+|) = 
\frac 14 (1+\sqrt{1-\gamma })^2 |\Phi ^+ \rangle \langle \Phi ^+|  + 
\frac 14 (1- \sqrt{1-\gamma })^2 |\Phi ^- \rangle \langle \Phi ^-|  
\nonumber \\& & 
+\frac \gamma 4 (|\Phi ^+ \rangle \langle \Phi ^-|+|\Phi ^- \rangle 
\langle \Phi ^+|+ |\Psi ^+ \rangle \langle \Psi ^+| + |\Psi ^- \rangle \langle \Psi ^-|
+|\Psi ^+ \rangle \langle \Psi ^-| + |\Psi ^- \rangle \langle \Psi ^+|)
\;.\label{outdamp}
\end{eqnarray}

\begin{figure}[htb]
  \includegraphics[scale=0.8]{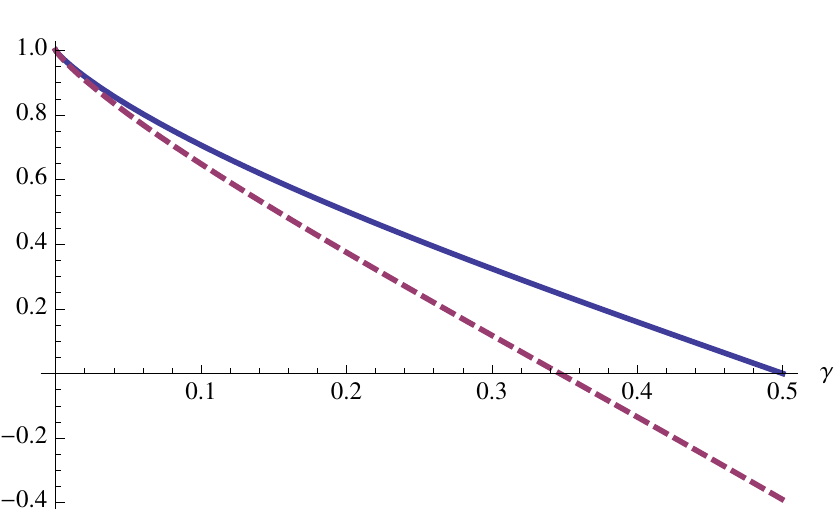}
  \caption{Amplitude damping channel with parameter $\gamma $: 
detected quantum capacity with maximally entangled input and estimation of $\vec p$ for the 
eigenstates of (\ref{outdamp}) and for the Bell basis
(solid and dashed line, respectively).} 
  \label{fig:damp2}
\end{figure}

The reduced output state corresponding to the system qubit alone 
is given by ${\cal E}\left (\frac I2\right )= 
\frac 12 (I +\gamma \sigma _z)$, 
hence  it has von Neumann entropy 
$S\left [{\cal E}\left (\frac Id \right )\right ]
=H_2 \left(\frac {1- \gamma }{2}\right )$. 
By performing the local measurement $ {\cal M}_2$
and optimising $\vec p$ over the bases 
(\ref{b1}-\ref{b3}) as in Eq. (\ref{opt}), one can detect the bound 
\begin{eqnarray}
Q\geq Q_{DET}&=& H_2 \left(\frac {1- \gamma }{2}\right )- H(\vec p)  =   
H_2 \left(\frac {1- \gamma }{2}\right )- H_2 \left(\frac \gamma  2 \right ) 
\,,\label{bounddamp}
\end{eqnarray}
where the optimal vector of probabilities is given by
 $\vec{p}= \left (1 -\gamma  /2 \,,0 \,,\gamma /2 \,, 0 \right )$,
and it corresponds to the basis in Eq. (\ref{b1}), with 
$a=\frac{1+\sqrt {1 - \gamma   }}
{\sqrt{2(2- \gamma )}}$, 
$b=\frac{\gamma }{(1+\sqrt {1 - \gamma   }) \sqrt{2(2- \gamma )}}$, 
and $c=d=\frac{1}{\sqrt 2}$.  In fact, 
this basis corresponds to the projectors on the eigenstates of the output 
state (\ref{outdamp}).  
It turns out that, as long as $\gamma < 1/2$, 
a non-vanishing quantum capacity is then detected. 
Indeed the difference $Q-Q_{DET}$ never exceeds $0.005$.  
We notice that for this form of noise the Bell basis (\ref{phipsi}) 
does not provide the minimum value of $H(\vec p)$. Actually, 
for the Bell basis one has 
\begin{eqnarray}
\vec{p}= \frac 14 \left (
(1 +\sqrt{1- \gamma } )^2 \,,(1  -\sqrt{1- \gamma  })^2 \,,\gamma  \,, \gamma \right )
\;,\label{bellvec}
\end{eqnarray}
and by using this value of $\vec p$ a non-vanishing quantum capacity is detected only for 
$\gamma < 0.3466 $. In Fig. \ref{fig:damp2} we plot the detectable 
bound from Eq. (\ref{bounddamp}) [which is indistinguishable from the quantum capacity (\ref{qcdamp})], 
along with the bound obtained by the probability vector (\ref{bellvec}) pertaining to the Bell projectors,  
versus the damping parameter $\gamma  $. The difference of the curves shows how
the optimisation of $Q_{DET}$ over the bases (\ref{b1}-\ref{b3}) is crucial to 
achieve the optimal bound. 

We want to point out that 
the above scheme has been successfully tested
experimentally for various models of qubit noise, including the phase damping,
depolarising and amplitude damping channels described above, by using pairs of 
polarised photons \cite{qcdet-exp}. 
We also want to stress that the detection method reviewed here
can be applied also to multipartite quantum channels
and it has proved to give a satisfactory theoretical performance 
in the case of two-qubit
correlated channels in Ref. \cite{qcdet-corr}.

We finally want to point out that all detectable bounds we are providing
also give lower bounds to the private information $P$,  
since $P \geq Q_1$ \cite{NC00}. 
Moreover, we can also derive a detectable lower bound to the 
entanglement-assisted classical capacity. 
Actually, the latter is defined as \cite{CE}
\begin{eqnarray} 
C_E = \max_{\rho } I(\rho , {\cal E}_1)
\;,\label{C_E} 
\end{eqnarray} 
where 
$I(\rho , {\cal E}_1) = S(\rho ) + I_c
(\rho, {\cal E}_1)$. 
By considering the procedure outlined above we then have the lower bound
\begin{eqnarray} 
C_E \geq \log_2 d+ Q_{DET}\;,
\label{C_E-lb} 
\end{eqnarray} 
where a maximally entangled
state $|\Psi \rangle$ is considered as input, giving $S(\rho )=\log_2 d$.

\section{Conclusions}
\label{s:conc}

In conclusion, we have reviewed in a unified way a method to detect properties
based on convexity features of quantum channels, and more generally of quantum
evolutions of open systems, and lower 
bounds to capacities of quantum communication channels, specifically to the 
quantum capacity, the entanglement 
assisted capacity, and the private capacity. 
The procedures we presented do not require any
a priori knowledge about the quantum channel and rely on a number of 
measurement 
settings that scales as $d^2$. They are therefore much cheaper than full 
process tomography, whose number of measurements in the entanglement based 
scenario considered here scales as $d^4$. Moreover, they can be easily
implementable in an experimental scenario without posing any 
particular restriction on the nature of the physical system under 
consideration. 
As shown in Sect. \ref{s:preli}, the method can be equivalently applied 
by suitably preparing the input and measuring the output system alone 
without necessarily requiring the use of entangled states.
We also point out that the same scheme outlined in Sect. 
\ref{s:preli} can also be applied to detect non-Markovianity properties for 
some classes of dynamical maps in open quantum systems \cite{noma-wit}.

\end{document}